\documentclass[pra, a4paper, amsfonts, amssymb, amsmath, notitlepage]{revtex4-2}
\usepackage[english]{babel}
\usepackage[utf8]{inputenc}
\usepackage{bbold}
\usepackage{amsthm}
\usepackage{mathtools}
\usepackage{physics}
\usepackage{xcolor}
\usepackage{graphicx}
\usepackage[left=23mm,right=13mm,top=35mm,columnsep=15pt]{geometry} 
\usepackage{adjustbox}
\usepackage{placeins}
\usepackage[T1]{fontenc}
\usepackage{lipsum}
\usepackage{csquotes}
\usepackage{soul,xcolor}

\renewcommand{\a}{{\hat{a}}}
\newcommand{\AAA}{{\mathbb{A}}}
\newcommand{\ad}{{\hat{a}^{\dagger}}}
\newcommand{\BBB}{{\mathbb{B}}}
\newcommand{\DD}{\hat{\hat{\mathcal{D}}}}
\newcommand{\HH}{\hat{\mathcal{H}}}
\newcommand{\FF}{\mathcal{F}}

\newcommand{\III}{{\mathbb{I}}}
\newcommand{\LL}{\hat{\hat{\mathcal{L}}}}
\newcommand{\Laplace}{\mathfrak{L}}
\newcommand{\lcav}{l_{\rm cav}}
\newcommand{\RR}{\mathcal{R}}

\newcommand{\XXX}{{\mathbb{X}}}

\newcommand{\sig}{{\hat{\sigma}}}

\begin{document}

\title{Ultimate stability of active optical frequency standards}

\author{Georgy A. Kazakov, Swadheen Dubey}
    \affiliation{Atominstitut, TU Wien, Stadionallee 2, 1020 Vienna, Austria}
\author{Anna Bychek}
	\affiliation{Institut f\"ur Theoretische Physik, Universit\"at Innsbruck,
	Technikerstr. 21a, A-6020 Innsbruck, Austria}
\author{Uwe Sterr}
    \affiliation{Physikalisch-Technische Bundesanstalt (PTB), Bundesallee 100, 38116 Braunschweig, Germany}
\author{Marcin Bober, Micha\l{}~Zawada}
    \affiliation{Institute of Physics, Faculty of Physics, Astronomy and Informatics, Nicolaus Copernicus University, Grudzi\c{a}dzka 5, PL-87-100 Toru\'n, Poland}
\date{\today} 

\begin{abstract}
Active optical frequency standards provide interesting alternatives to their passive counterparts. 
Particularly, such a clock alone continuously generates highly-stable narrow-line laser radiation. Thus a local oscillator is not required to keep the optical phase during a dead time between interrogations as in passive clocks, but only to boost the active clock's low output power to practically usable levels with the current state of technology. 
Here we investigate the spectral properties and the stability of active clocks, including homogeneous and inhomogeneous broadening effects. 
We find that for short averaging times the stability is limited by photon shot noise from the limited emitted laser power and at long averaging times by phase diffusion of the laser output. 
Operational parameters for best long-term stability were identified. 
Using realistic numbers for an active clock with $^{87}$Sr we find that an optimized stability of 
$\sigma_y(\tau) \approx 4\times10^{-18}/\sqrt{\tau [\mathrm{s}]}$ is achievable.

\end{abstract}

\keywords{active optical clocks, superradiance}

\maketitle
\section{Introduction}
\label{sec:intro}

Modern-day optical clocks are passive frequency standards~\cite{Ludlow_2015}, where the frequency of a laser pre-stabilized to an ultra-stable optical cavity is periodically compared with the frequency of a narrow and robust {\em clock transition} in a sample of trapped atoms (or ions). 
The measurement sequence includes an interrogation time, during which the phase of the laser is imprinted to the atomic sample, and a {\em dead time}, when the laser pre-stabilised to an ultra-stable macroscopic cavity keeps the frequency, playing the role of a flywheel. 
Such a clock has demonstrated an excellent stability at the level of $6.6\times10^{-19}$ after 1 hour of averaging \cite{Oelker_2019}, however, on shorter timescale this stability is limited by thermal and mechanical fluctuations of the length of this ultra-stable cavity. 
This problem may be overcome with the help of an {\em active optical frequency standard} based on a laser operating deep in the bad-cavity regime \cite{Meiser_2009, Chen_2009}, where the linewidth of the cavity is much broader than the linewidth of the gain. 
The gain of such a laser can be formed by forbidden transitions in alkaline-earth atoms, the same as used for passive optical lattice clocks. 
Similar to a hydrogen maser, the frequency of such a laser is determined by the frequency of lasing transition and is robust to fluctuations of the cavity length, which improves the stability on shorter timescales. 

In the present paper we study the stability that can be attained with such a laser and compare it with the one of a passive optical clock based on an atomic ensemble with similar characteristics. 
For the sake of definiteness, we consider the model of two-level laser with continuous incoherent repumping \cite{Meiser_2009}.
Bad-cavity lasers based on other schemes, such as atomic beam lasers \cite{Chen_2009}, optical conveyor lasers \cite{Kazakov_14}, and lasers with sequential coupling of atomic ensembles \cite{Kazakov_13} should have similar characteristics, up to some numerical factors. 
In Section~\ref{sec:AOFCstab} we present general expressions for the short-term stability of a secondary laser phase locked to a low-power narrow-line continuous-wave bad-cavity laser.
In Section~\ref{sec:linewidth} we calculate the linewidth of the bad-cavity laser's Lorentzian spectrum and discuss how this linewidth depends on the natural linewidth of the lasing transition in the employed gain atoms, on inhomogeneous broadening and dephasing of the atomic transition, on the number of atoms providing the gain, and on parameters of the cavity. 
We optimize the cooperativity as well as the rate of incoherent pumping to attain a minimum linewidth at a given atomic number and cavity finesse. We express these optimized parameters as well as the linewidth and the respective number of intracavity photons via characteristic properties of the atomic ensemble.
In Section~\ref{sec:estimation} we estimate the achievable performance for ensembles of atoms trapped in an optical lattice potential and compare the respective frequency stabilities that can be obtained with the help of active and passive frequency standards based on such ensembles.

\section{Active optical frequency standard and its stability}
\label{sec:AOFCstab}
The spectral characteristics of a bad cavity laser's output field $E$ can be described by its power spectral density $S_E(f)$. 
It can be obtained from the two-time correlation function $\RR(\tau)$ with the help of the well-known Wiener-Khinchin theorem, see equation~(\ref{eq:21}) in Section~\ref{Sec::modelCalc} and \cite{Debnath18}. 
In first approximation $\RR(\tau)$ may be described by an exponentially decaying function that corresponds to a Lorentzian lineshape of $S_E(f)$ centered at an ordinary frequency $f_0=\omega_0/(2\pi)$ with half-width $\Delta f = \Delta \omega/ (2 \pi)$.
Such a signal has white frequency noise with a single sided spectral power density $S_y(f)$ of fractional frequency fluctuations $y=\Delta \omega / \omega_0$ equal to
\begin{equation}
    S_y(f) = \frac{\Delta f}{\pi f_0^2}=\frac{2 \Delta \omega}{\omega_0^2},
    \label{eq:1}
\end{equation}
corresponding to a spectral power density $S_\phi(f)$ of phase fluctuations
\begin{equation}
    S_\phi(f) = \frac{\Delta f}{\pi f^2} = {2 \Delta \omega}{f^2},
    \label{eq:2}
\end{equation}
and Allan deviation 
\begin{equation}
	\sigma^\prime_y(\tau) 
	= \sqrt{ \frac{\Delta \omega}{\omega_0^2 \tau}}.
	\label{eq:3}
\end{equation}

In addition, due to the finite rate of emitted photons, the field of power $P$ shows quantum fluctuations, leading to a limited signal-to-noise ratio expressed as the ratio of signal power to power of the noise per unit bandwidth 
$\mathrm{SNR} = P/(\hbar\omega_0)$ \cite{Teich72}.
Theses fluctuations appear as white amplitude and phase noise of the signal. 
When the active-laser output is heterodyned with an ideal powerful and perfectly stable cw laser, the amplitude noise is usually of no importance to the frequency stability, and the power spectral density of white phase noise $S_\phi$ amounts to 
\begin{equation}
	S_\phi(f) = \mathrm{SNR}^{-1}=\frac{\hbar \omega_0}{P},
	\label{eq:4}
\end{equation}
with the corresponding Allan deviation \cite{rub05, daw07}
\begin{equation}
    {\sigma^{\prime\prime}_y(\tau) = \frac{1}{\tau} \sqrt{\frac{3 \hbar f_h}{\omega_0 P}}}.
    \label{eq:5}
\end{equation}
As the Allan deviation would diverge for white phase noise with unlimited bandwidth, the noise is set to zero for frequencies above a cut-off frequency $f_h$ (in ordinary frequency units) to obtain a finite value.
In practice this low-pass behavior can appear from the bandwidth of a phase locked loop using the heterodyne signal.

To avoid the dependence on the arbitrary cut-off frequency, in this case the modified Allan deviation is often used:
\begin{equation}
    {\text{mod}\, \sigma^{\prime\prime}_y(\tau) = \frac{1}{\tau^{3/2}} \sqrt{\frac{3 \hbar}{2 \omega_0 P}}}.
    \label{eq:6}
\end{equation}
Adding the random walk noise of the phase associated with damping of two-time correlation of the cavity field and the white phase noise associated with shot noise in the number of emitted photons results in the overall Allan deviation
\begin{equation}
    { \sigma_y(\tau) = 
         \sqrt{(\sigma^{\prime}_y(\tau))^2
       +(\sigma^{\prime\prime}_y(\tau))^2} 
    = 
       \sqrt{\frac{\Delta \omega}{\omega_0^2 \tau} 
       +\frac{3 \hbar f_h}{\omega_0 P \tau^2}}}.
    \label{eq:7}
\end{equation}
and the overall modified Allan deviation
\begin{equation}
    {\text{mod}\, \sigma_y(\tau) = 
     \sqrt{\text{mod}\,  \sigma^{\prime}_y(\tau)^2+\text{mod}\, \sigma^{\prime\prime}_y(\tau)^2} 
    = \sqrt{\frac{\Delta \omega}{2 \omega_0^2 \tau} 
     +\frac{3 \hbar}{2 \omega_0 P \tau^3}}}.
    \label{eq:8}
\end{equation}

\noindent
At short averaging times $\tau$ it is determined by the bad-cavity laser's output power $P$ and at long times by its linewidth $\Delta \omega$.

The contribution $\sigma^{\prime\prime}_y(\tau)$ (\ref{eq:5}) to the total instability $\sigma_y(\tau)$ is associated with the photon shot noise. Its influence depends on the bandwidth of the feedback loop to phase lock a secondary laser with good short-term stability to the bad cavity laser (see discussion in Section~\ref{sec:estimation}).
The contribution $\sigma^{\prime}_y(\tau)$ (\ref{eq:3}) is more fundamental in that sense that it does not depend on the properties of the secondary laser and it limits the stability on longer timescale.

In the next section we consider a generic model of a two-level bad cavity laser with incoherent pumping and find general expressions for the minimum linewidth $\Delta \omega$ and the necessary set of optimized parameters.

\section{Linewidth of a bad cavity laser}
\label{sec:linewidth}

In this section we overview the dependence of the linewidth on the characteristics of the bad-cavity laser with continuous incoherent repumping and estimate the minimum linewidth which can be achieved in such type of laser. First we consider a two-level model of a bad-cavity laser with incoherent pumping, as studied in \cite{Meiser_2009}. Such a laser has two lasing thresholds $R_{\rm min}$ and $R_{\rm max}$; below the lower threshold $R_{\rm min}$ the pumping is not enough to create the necessary inversion for the lasing and above the upper threshold $R_{\rm max}$ the pumping destroys the coherence, thus also preventing the coherent emission. 
In the homogeneous case (i.e., when all atoms contributing to the gain have exactly the same parameters, such as coupling strength with the cavity field, transition frequency, dephasing rate, etc.), and when the laser operates far from the lower and the upper lasing thresholds, the linewidth $\Delta \omega_\text{min}$ of such a laser can be estimated~\cite{Meiser_2009} as 
\begin{equation}
\Delta \omega_\text{min} \approx C \gamma_s = 4 g^2/\kappa.
\label{eq:9}
\end{equation}
\noindent
 Here $\kappa$ is the decay rate of the energy of the cavity field, $g$ is the coupling strength between the laser field and the atomic transition (the Hamiltonian is presented in expression~\ref{eq:11}), $\gamma_s$ is the spontaneous rate of the lasing transition, and $C=4g^2/(\kappa\gamma_s)$ is the {\em cooperativity parameter}. 
 It may seem that one should just take the cooperativity $C$ as small as possible to minimize the linewidth. However, expression~(\ref{eq:9}) is valid only if the pumping rate $R$ is much bigger than the lower and much smaller than the upper lasing thresholds $R_{\rm min}$ and $R_{\rm max}$ respectively. 
 Accurate expressions for these thresholds in the homogeneous case will be derived in section~\ref{Sec::modelHomogen}. 
 One may see from expressions (\ref{eq:32}) and (\ref{eq:33}) that both lasing thresholds approache each other when the cooperativity $C$ decreases at a given number $N$ of atoms.
 Therefore, a minimum linewidth is attained in such a range of parameters where the condition $R_{\rm min} \ll R \ll R_{\rm max}$ is not fulfilled anymore and where the estimate~(\ref{eq:9}) is not valid. Thus we need to find a more accurate estimate for $\Delta \omega_\text{min}$.
 
The spectral properties of a continuous-wave laser can be derived from the two-time correlation function of its output field $\RR(\tau)=\langle \ad(t_0+\tau)\a(t_0)\rangle$, which in the bad-cavity regime is directly proportional to the correlation of the atomic coherence \cite{Meiser_2009}. 
In the present paper we limit our consideration to a model where the laser gain is formed by $N$ two-level atoms subjected to incoherent pumping and coupled to the cavity field. 
Such a two-level model can correctly represent the dynamic of a real multilevel superradiant laser with continuous repumping and single lasing transition, if the lifetimes of the intermediate levels are much shorter than any other timescale in the system except, may be, the decay rate of the cavity field \cite{Hotter21}. 
Because the Hilbert space describing such a system grows exponentially with atom number $N$, one has to use some approximation to reduce the problem size. 
We restrict our consideration to a second-order cumulant approximation, following \cite{Meiser_2009} and \cite{Bychek2021}, which allows calculating both output power and spectrum of the superradiant laser. 
In subsection~\ref{Sec::modelInhomogen} we briefly overview the model and explain the most important details of the calculation. 
In subsection~\ref{Sec::modelHomogen} we consider the particular case of a homogeneous system, where all the atoms are equally coupled to the cavity field and share the same transition frequency and all other parameters. 
We obtain analytical expressions for the output power and the linewidth in this simplest case and perform a qualitative analysis of their dependencies. 
In subsection~\ref{Sec::modelCalc} we study the linewidth quantitatively, both for the simple homogeneous model and for a more realistic model with inhomogeneous coupling of the atoms to the cavity field and inhomogeneous broadening of the lasing transition.

\begin{figure}
    \centering
    \includegraphics[width=0.2\textwidth]{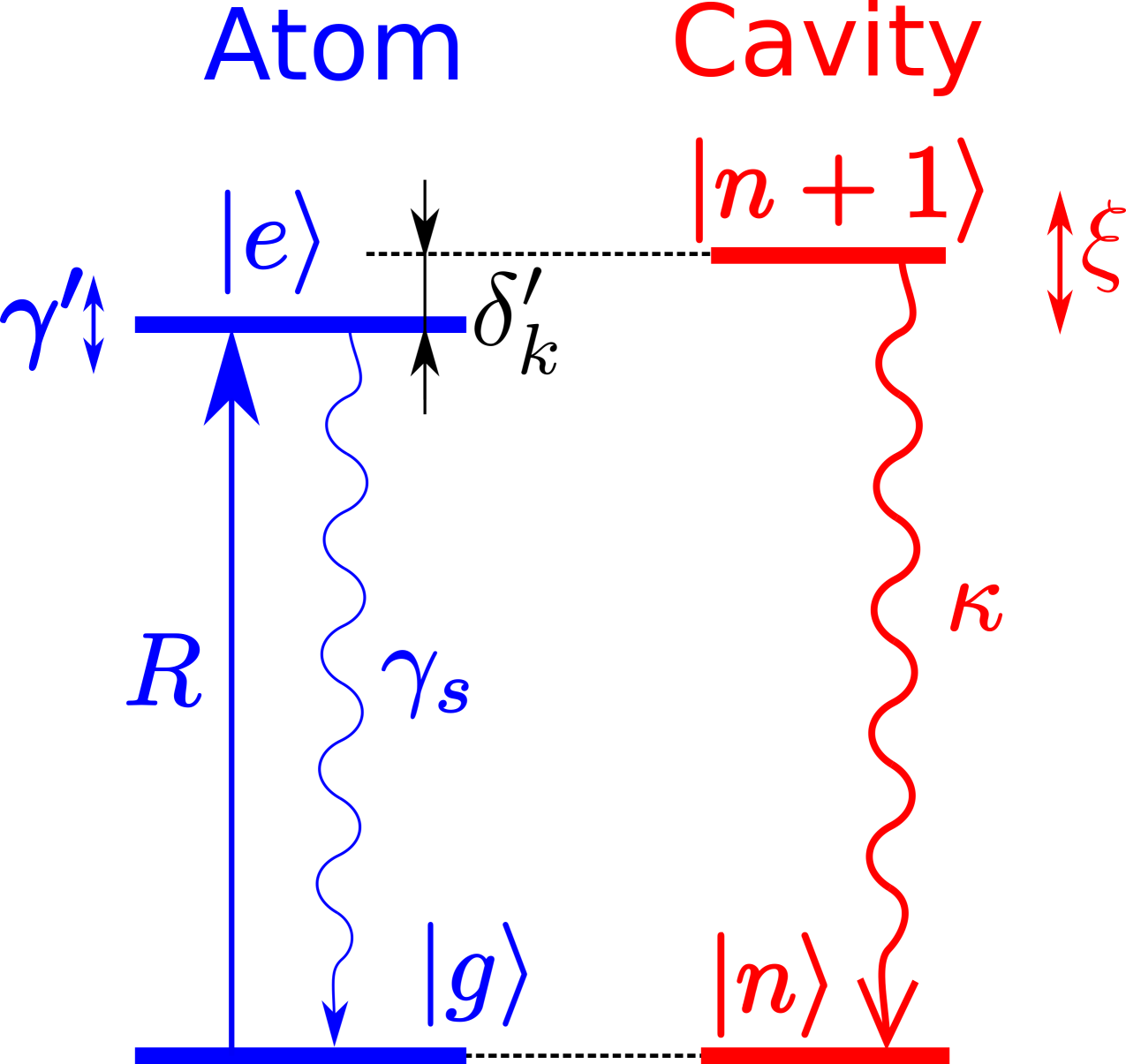}
    \caption{Level diagram of a 2-level atom coupled to the cavity field. 
    Here $\delta'_k$ is the shift between the cavity and the atomic transition frequency, 
    $R$ is the incoherent pumping rate, 
    $\gamma_s$ is the spontaneous decay rate, 
    $\gamma'$ is the dephasing rate of the atom, 
    $\kappa$ is the energy decay rate of the cavity field, and $\xi$ is the cavity dephasing rate.}
    \label{fig:2level}
\end{figure}
\subsection{Inhomogeneous system: description of the model and equations}
\label{Sec::modelInhomogen}
We consider an ensemble of $N$ two-level atoms confined in space (for example, with the help of an optical lattice potential) and interacting with a single cavity mode, see Figure~\ref{fig:2level}. We neglect dipole-dipole interactions between different atoms as well as collective coupling of the atoms to the bath. The averaged value of an operator $\hat{O}$ describing such a system can be written as
\begin{equation}
\frac{\langle d\hat{O} \rangle}{dt} = \frac{i}{\hbar} \langle[\HH,\hat{O}] \rangle+\langle \LL[\hat{O}] \rangle.
    \label{eq:10}
\end{equation}
The Hamiltonian $\HH$ in the rotating frame can be written as
\begin{equation}
    \HH = \hbar \left[ \delta_c \ad \a + \sum_{j=1}^N g_j (\sig_{eg}^j \a + \ad \sig_{ge}^j) + \sum_j \Delta_j \sig_{ee}^j\right],
    \label{eq:11}
\end{equation}
were $\ad$ and $\a$ are field creation and annihilation operators, index $j$ runs over the atoms, 
$\sig_{\alpha \beta}^j=|\alpha^j\rangle \langle \beta^j | \bigotimes_{k\neq j} \hat{\mathbb{1}}^k$ are single-atom transition operators, 
$|\alpha^j \rangle $ and $|\beta^j\rangle$ run over ground $|g^j\rangle$ and excited $|e^j\rangle$ states of $j$th atom, 
$g_j$ is a coupling coefficient between $j$th atom and the field,
$\Delta_j$ is the shift of the transition in $j$th atom caused by some non-homogeneous effects, and $\delta_c$ is the shift of the cavity resonance frequency from the frequency of our rotating frame.

The Liouvillian term describing the dissipative process is equal to
\begin{equation}
    \LL[\hat{O}] = \kappa \DD[\a] \hat{O} + \xi \DD[\ad \a] \hat{O} 
    + \sum_{j=1}^N \left[ \gamma_s \DD[\sig_{ge}^j] \hat{O} 
        + R_j \DD[\sig_{eg}^j] \hat{O} + \gamma'_j \DD[\sig_{ee}^j] \hat{O} \right],
    \label{eq:12}
\end{equation}
where $\DD[\hat{p}]\hat{O} =  \hat{p}^\dagger \hat{O} \hat{p} -\frac{1}{2}(\hat{p}^\dagger \hat{p} \hat{O} +\hat{O} \hat{p}^\dagger \hat{p})$ is a Lindbladian superoperator. 
Here $\kappa$ is the decay rate of the energy of the cavity mode, $\gamma_s$ is the spontaneous decay rate of the upper lasing state, $\xi$ is the dephasing rate of the cavity field, $R_j$ and $\gamma'_j$ are the rates of incoherent pumping and dephasing of $j$th atom.

The closed set of differential equations for the stochastic means of system operators can be written with the help of a cumulant expansion up to the second order and the phase invariance as in \cite{Bychek2021}
\begin{align}
\frac{d}{dt}\langle \ad \a \rangle &= - \kappa \langle \ad \a \rangle 
    + i \sum_{j=1}^N g_j (\langle \sig_{eg}^j \a \rangle - \langle \ad \sig_{ge}^j \rangle),
\label{eq:13} 
\\
\frac{d}{dt}\langle \sig_{eg}^k \a \rangle &
    = - \left[ \frac{\kappa'_k}{2} + i \delta'_k \right] \sig_{eg}^k 
      + i g_k \left[\langle \ad \a \rangle(1-2\langle \sig_{ee}^k \rangle) 
      - \langle \sig_{ee}^k \rangle \right] 
      - i \sum_{j\neq k} g_j \langle \sig_{eg}^k \sig_{ge}^j \rangle,
\label{eq:14} 
\\
\frac{d}{dt}\langle \sig_{ee}^k \rangle &
    = i g_k \left[ \langle \ad \sig_{ge}^k \rangle - \langle \sig_{eg}^k \a \rangle \right]
      - (\gamma_s+R_k) \langle \sig_{ee}^k \rangle + R_k,
\label{eq:15} 
\\
\frac{d}{dt}\langle \sig_{eg}^k  \sig_{ge}^l \rangle &
    = - \left[ \Gamma'_{kl}+i \Delta_{lk}\right] \langle \sig_{eg}^k  \sig_{ge}^l \rangle
      - i g_k \langle \ad \sig_{ge}^l \rangle (2 \langle \sig_{ee}^k \rangle-1) 
      + i g_l \langle \sig_{eg}^k \a \rangle (2 \langle \sig_{ee}^l \rangle-1),
\label{eq:16} 
\end{align}
where $\kappa'_k=\kappa+\xi+R_k+\gamma'_k+\gamma_s$, $\delta'_k=\delta_c-\Delta_k$, $\Gamma'_{kl}=\gamma_s+(R_k+R_l+\gamma'_k+\gamma'_l)/2$ and $\Delta_{lk}=\Delta_l-\Delta_k$. 
These equations can, in principle, be solved numerically. However, the number of equations scales quadratically with the number of the atoms. 
For practical simulations of ensembles with tens of thousands of atoms, one needs to group the atoms into $M$ clusters, where all $N_j$ atoms of $j$th cluster are considered as identical. 
Also, if the rate $\kappa$ is much larger than all the evolution rates of atomic polarizabilities, it is convenient to perform an adiabatic elimination of the fast variables $\langle \ad \a \rangle$, $\langle \ad \sig_{ge} \rangle$, $\langle \sig_{eg} \a \rangle$. 
Then one may express
\begin{align}
\langle \ad \a \rangle & = 
    \left[ 
        \kappa - \sum_k \frac{4 N_k g_k^2 \kappa'_k}{\kappa'^2_k+4 \delta'^2_k}
        [2\langle \sig_{ee}^k\rangle - 1]
    \right]^{-1}
\nonumber \\
&  \times
    \sum_{k} 
    \frac{4 g_k N_k}{\kappa'^2_k+4 \delta'^2_k}
    \left[ \kappa'_k 
        \left( g_k \langle \sig_{ee}^k \rangle 
            +\sum_j N'_{j,k} g_j \Re(\langle \sig_{eg}^k \sig_{ge}^j \rangle) 
        \right)
    \right.
 \left. + 2 \delta'_k
         \left(
            \sum_j N'_{j,k} g_j \Im(\langle \sig_{eg}^k \sig_{ge}^j \rangle)
         \right)
    \right]
\label{eq:17}
\end{align}
and 
\begin{align}
\langle \sig_{eg}^k \a \rangle & = 
    \frac{2}{\kappa'_k+2 i \delta'_k} 
    \left\{ig_k 
        \left[ \langle \ad \a \rangle(1-2 \langle \sig_{ee}^k \rangle) - \langle \sig_{ee}^k \rangle
        \right]
        -i\sum_j N'_{j,k} g_j \langle \sig_{eg}^k \sig_{ge}^j \rangle
    \right\}.
\label{eq:18}
\end{align}
Here the sums are taken over clusters instead of atoms, $N_k$ is the number of atoms in the cluster, and 
\begin{equation}
    N'_{j,k}=\left\{ 
    \begin{array}{lc}
         N_j, & j\neq k  \\
         \max(0,N_k-1),&  j=k
    \end{array}
    \right.
\label{eq:19}
\end{equation}
Substituting expressions (\ref{eq:17}) and (\ref{eq:18}) into equations (\ref{eq:15}) and (\ref{eq:16}), and solving them numerically, one can find the steady-state values of $\langle \sig^j_{eg} \a\rangle$ and $\langle \sig^j_{ee} \rangle$, if only the atomic dipoles get synchronized. Then one may express the steady-state values of $\langle \ad \a \rangle$, $\langle \sig_{eg}^j \a \rangle$ and $\langle \ad \sig_{ge}^j \rangle$ with the help of equations (\ref{eq:17}) and (\ref{eq:18}). The output power $P$ of the laser is equal to
\begin{equation}
P = \eta \hbar \omega_0 \kappa \langle \ad \a \rangle,
\label{eq:20}
\end{equation}
where $\eta$ is the relative transmission of the outcoupling mirror, and $\omega_0$ is the angular frequency of the laser radiation. 

Finally, let us discuss how to calculate the spectrum of the superradiant laser. 
According to the Wiener-Khinchin theorem, the spectral density $S_E(f)$ of the signal can be obtained as a real part of Fourier transform of the 2-time correlation function $\RR(\tau)=\langle \ad(t_0+\tau)\a(t_0)\rangle$:
\begin{equation}
S_E(f) \propto \text{Re}\int\limits_0^\infty \RR(\tau) e^{-2\pi i f \tau} d\tau.
\label{eq:21}
\end{equation}
In an established steady-state regime $\langle \ad(t_0+\tau)\a(t_0)\rangle=\langle \ad(\tau)\a(0)\rangle \equiv \langle \ad\a_0\rangle$, where $\ad=\ad(t)$, and $\a_0=\a(0)$. To find this function, one needs to solve the set of equations obtained with the help of the quantum regression theorem
\begin{align}
\frac{d}{dt}\langle \ad \a_0 \rangle &= 
	-\left[
		\frac{\kappa + \xi}{2} -i\delta_c
	\right] \langle \ad \a_0 \rangle
	+i\sum_k N_k g_k \left\langle \sig^k_{eg} \a_0 \right\rangle,
\label{eq:22}
\\
\frac{d}{dt}\langle \sig^k_{eg} \a_0 \rangle &= 
	-\left[
		\frac{\gamma_s+R_k+\gamma'_k}{2} -i\Delta_k
	\right] \langle \sig^k_{eg} \a_0 \rangle
	-i g_k \left\langle \sig^k_{z}\right\rangle
	\langle \ad \a_0 \rangle.
\label{eq:23}
\end{align}
where $\left\langle \sig^k_{z}\right\rangle=\left\langle \sig^k_{ee}\right\rangle-\left\langle \sig^k_{gg}\right\rangle $. Substituting here the established time-independent values of $\langle \sig^k_{z}\rangle$ into (\ref{eq:22}) and (\ref{eq:23}) and performing the Laplace transform, one obtains a set of linear equations of the form $(\AAA + \III\, s) \cdot \XXX = \BBB$, where $\III$ is identity matrix,
\begin{align}
\AAA &= 
	\left[ 
		\begin{array}{cccc}
			\frac{\kappa + \xi}{2}-i \delta_c 
				& -i N_1 g_1 
				& \cdots ~ \cdots
				& -i N_M g_M 
			\\
			ig_1 \left\langle \sigma_z^1 \right\rangle 
				& \frac{\gamma+R_1+\gamma'_1}{2}-i\Delta_1
				& \cdots 0 \cdots & 0 
			\\		
			\vdots & \vdots & \ddots & \vdots \\
			ig_M \left\langle \sigma_z^M \right\rangle 
				& 0
				& \cdots 0 \cdots 
				& \frac{\gamma+R_M+\gamma'_M}{2}-i\Delta_M 			
		\end{array}
	\right], 
&
\BBB &=
	\left[
		\begin{array}{c}
			\langle \ad \a \rangle_{s} \\
			\langle \sig^1_{eg} \a \rangle_{s} \\
			\vdots \\
			\langle \sig^M_{eg} \a \rangle_{s}
		\end{array}
	\right],
&
\XXX &=
	\left[
		\begin{array}{c}
			\Laplace\{\langle \ad \a_0 \rangle\}(s) \\
			\Laplace\{\langle \sig^1_{eg} \a_0 \rangle\}(s) \\
			\vdots \\
			\Laplace\{\langle \sig^M_{eg} \a_0 \rangle\}(s)
		\end{array}
	\right],
\label{eq:24}
\end{align}
$\Laplace\{f\}(s)=\int_0^\infty f(t) e^{-st}dt$ is the Laplace transform, and the subscript $s$ denotes ``steady-state''. 
Using the connection between Laplace and Fourier transforms, one can calculate the power spectral density of the bad-cavity laser output
\begin{equation}
S_E(f) \propto \text{Re} \left[\Laplace\{\langle \ad \a_0 \rangle\}(2 \pi i f)\right].
\label{eq:25}
\end{equation}

From the power spectral density obtained with the help of (\ref{eq:24}) and (\ref{eq:25}), one obtains the lasers's full linewidth at half maximum $\Delta f = \Delta \omega / (2 \pi)$.

\subsection{Homogeneous case: analytic expressions and qualitative considerations}
\label{Sec::modelHomogen}
In this section we consider the simplest case of a bad-cavity laser with homogeneous gain, i.e., the situation when all the atoms have the same transition frequency $\omega_a$, pumping and dephasing rate $R$ and $\gamma'$ and coupling strength $g$ to the cavity field.
The steady-state solution and the linewidth for such a simple system in second-order cumulant approximation can be found analytically or semi-analytically. 
This analysis has been partially performed, for example, in \cite{Meiser_2009}, and here we overview the main results and derive a few new useful relations. 
The correspondence between our notation and notation used there is the following: $R=w$, $\gamma_s=\gamma$, $\gamma'= 2/T_2$, and $g=\Omega/2$.

First, from equations (\ref{eq:13}) -- (\ref{eq:15}) one may easily express that in the homogeneous case
\begin{align}
\langle \ad \a \rangle_s &= \frac{N (\gamma_s+R)}{2 \kappa} \left(\frac{R-\gamma_s}{R+\gamma_s} - \langle \sig_z \rangle_s \right) 
\label{eq:26}\\
\langle \sig_{eg}^1 \sig_{ge}^2 \rangle_s &=\frac{ \langle \sig_z \rangle_s (\gamma_s+R) }{2\Gamma'} \left(\frac{R-\gamma_s}{R+\gamma_s} - \langle \sig_z \rangle_s \right), 
\label{eq:27}
\end{align}
where $\Gamma'=\gamma_s+R+\gamma'$, $\kappa'=\kappa+\xi+\Gamma'$.
Substituting these expressions into equation (\ref{eq:17}), one may obtain, after some algebra, the following quadratic equation for $\langle \sig_z\rangle_s$:
\begin{align}
\langle \sig_z \rangle_s^2 \left(\frac{N(\gamma_s+R)}{2\kappa}+\frac{(N-1)(\gamma_s+R)}{2 \Gamma'}\right) + \frac{(R-\gamma_s)(\kappa'^2+4(\delta_c-\Delta)^2)}{8 g^2 \kappa'} - \frac{1}{2} & 
\nonumber \\ 
- \langle \sig_z\rangle_s
\left[
    \frac{(R+\gamma_s)(\kappa'^2+4(\delta_c-\Delta)^2)}{8 g^2 \kappa'}
    +\frac{1}{2} + \frac{(R-\gamma_s)}{2}\left(\frac{N}{\kappa}+\frac{N-1}{\Gamma'} \right)
\right] &=0 .
\label{eq:28}
\end{align}
Solving this equation, we obtain the steady-state values $\langle \sig_z \rangle_s$, as well as $\langle \ad \a \rangle_s$ and $\langle \sig_{eg}^1 \sig_{ge}^2 \rangle_s$ with the help of (\ref{eq:26}) and (\ref{eq:27}).

Further in this Section we suppose, for the sake of simplicity, that all the atoms are in resonance with the cavity ($\delta_c=\Delta=0$), and that the cavity dephasing rate $\xi$ is negligible ($\xi = 0$). Then the equation (\ref{eq:28}) simplifies to
\begin{align}
\langle \sig_z \rangle_s^2 \left(\frac{N(\gamma_s+R)}{2\kappa}+\frac{(N-1)(\gamma_s+R)}{2 \Gamma'}\right) + \frac{(R-\gamma_s)\kappa'}{8 g^2} - \frac{1}{2} & 
\nonumber \\ 
- \langle \sig_z\rangle_s
\left[
    \frac{(R+\gamma_s)\kappa'}{8 g^2}
    +\frac{1}{2} + \frac{(R-\gamma_s)}{2}\left(\frac{N}{\kappa}+\frac{N-1}{\Gamma'} \right)
\right] &=0.
\label{eq:29}
\end{align}

Consider the equation (\ref{eq:29}). First, taking $N\approx N-1$ and neglecting $\gamma_s$, $R$ and $g$ in comparison with $\kappa$, one may express its approximate solutions as
\begin{equation}
\langle \sig_z \rangle_{s,1} \approx \frac{\kappa \Gamma'}{4 g^2 N},
\quad 
\langle \sig_z \rangle_{s,2} \approx \frac{R-\gamma_s}{R+\gamma_s},
\label{eq:30}
\end{equation}
and only the first solution gives $\langle \ad \a \rangle_s \neq 0$. This solution allows us to estimate the lasing thresholds. Substituting (\ref{eq:30}) into (\ref{eq:26}), one may find that lasing is possible, i.e., $\langle \ad \a \rangle_s > 0$, only if
\begin{align}
\frac{R-\gamma_s}{R+\gamma_s} > \frac{\kappa(\gamma_s+R+\gamma')}{4g^2 N} = \frac{\gamma_s+R+\gamma'}{NC\gamma_s},
\label{eq:31}
\end{align}
where we have introduced the cooperativity parameter $C = 4 g^2/(\kappa \gamma_s)$, in order to find limits of the pumping rate $R$:
\begin{equation}
\begin{split}
R_{\rm min} =\frac{NC\gamma_s-\gamma'-\sqrt{(NC\gamma_s-\gamma')^2-8\gamma_s^2 NC}}{2}-\gamma_s, \\
R_{\rm max} =\frac{NC\gamma_s-\gamma'+\sqrt{(NC\gamma_s-\gamma')^2-8\gamma_s^2 NC}}{2}-\gamma_s.
\end{split}
\label{eq:32}
\end{equation}
With $\gamma_s, \gamma' \ll NC\gamma_s$ it gives
\begin{equation}
\begin{split}
R_{\rm min} &\approx \gamma_s \frac{NC\gamma_s+\gamma'}{NC\gamma_s-\gamma'}, \\
R_{\rm max} &\approx NC\gamma_s-\gamma',
\end{split}
\label{eq:33}
\end{equation}
in correspondence with \cite{Kazakov2017}. 

The spectrum for the homogeneous case can be found from the set of linear equations (\ref{eq:22}) and (\ref{eq:23}) where, instead of performing the Laplace transform of the solution, we can just calculate $\Delta \omega $ as $ \Delta \omega = 2|\lambda|$. 
Here $\lambda$ is the eigenvalue with smallest absolute value of the matrix of this system (which can be easily proven by Fourier transform of exponentially decaying term in $\langle \ad \a_0 \rangle$). Taking $\kappa \gg |\lambda|$, one may express
\begin{align}
\Delta \omega = \Gamma'-\frac{4 g^2 N \langle \sig_z \rangle_s}{\kappa}.
\label{eq:34}
\end{align}
One may see that to calculate the linewidth one has to go beyond the semiclassical approximation: indeed, an attempt to substitute $\langle\sig_z\rangle_{s,1}$ from (\ref{eq:30}) into (\ref{eq:34}) gives $\Delta \omega = 0$. 
The straightforward way to calculate $\Delta \omega$ is to solve the quadratic equation (\ref{eq:29}) exactly, however, the result occurs to be too bulky for simple qualitative analysis. 
Instead, we calculate a correction to the approximate solution (\ref{eq:30}), expanding the coefficients of equation (\ref{eq:29}) into Fourier series.
After some algebra we get
\begin{align}
\Delta \omega \approx \frac{\Gamma'(\Gamma'+NC\gamma_s)}{2 \langle \ad \a \rangle_s \kappa} - \frac{\Gamma'}{N}.
\label{eq:35}
\end{align}
In the limit $\gamma_s, \gamma' \ll R \ll NC\gamma_s$ it gives $\Delta \omega \approx C\gamma_s$. This result has been reported in \cite{Meiser_2009} as a minimum attainable linewidth at given cooperativity $C$. We can not, however, take $C$ arbitrary small, otherwise we get into situation where $R_{\rm min}>R_{\rm max}$, and lasing becomes impossible. The minimum value of $C$, above which the lasing is still possible, can be found from equalizing $R_{\rm min}$ and $R_{\rm max}$ in (\ref{eq:32}), {which gives} 
\begin{align}
(NC_\text{min}\gamma_s - \gamma')^2 = 8 N C_\text{min} \gamma_s^2.
\label{eq:36}
\end{align}
At $\gamma'=0$ this minimum value is $C_\text{min} = 8/N$. Moreover, at very small $C$ the condition $\gamma_s, \gamma' \ll R \ll NC\gamma_s$ also can not be fulfilled, and the optimal value of $C$, where the minimum linewidth is attained, is larger than (but proportional to) $C_\text{min}$. 

We can conclude that the minimum attainable linewidth $\Delta \omega_\text{min}$ is proportional to $\gamma_s/N$. Therefore, it is convenient to express $\Delta \omega$ in units of $\gamma_s/N$ as a function of $CN$. Also, from expressions (\ref{eq:26}) and (\ref{eq:30}) we note that the dimensionless value $\langle \ad \a \rangle \kappa / (N \gamma_s)$ does not depends on $\kappa$ and $N$ at given values of $CN$, $R/\gamma_s$ and $\gamma'/\gamma_s$.

\begin{figure}[b]
    \centering
    \includegraphics[width=0.85\textwidth]{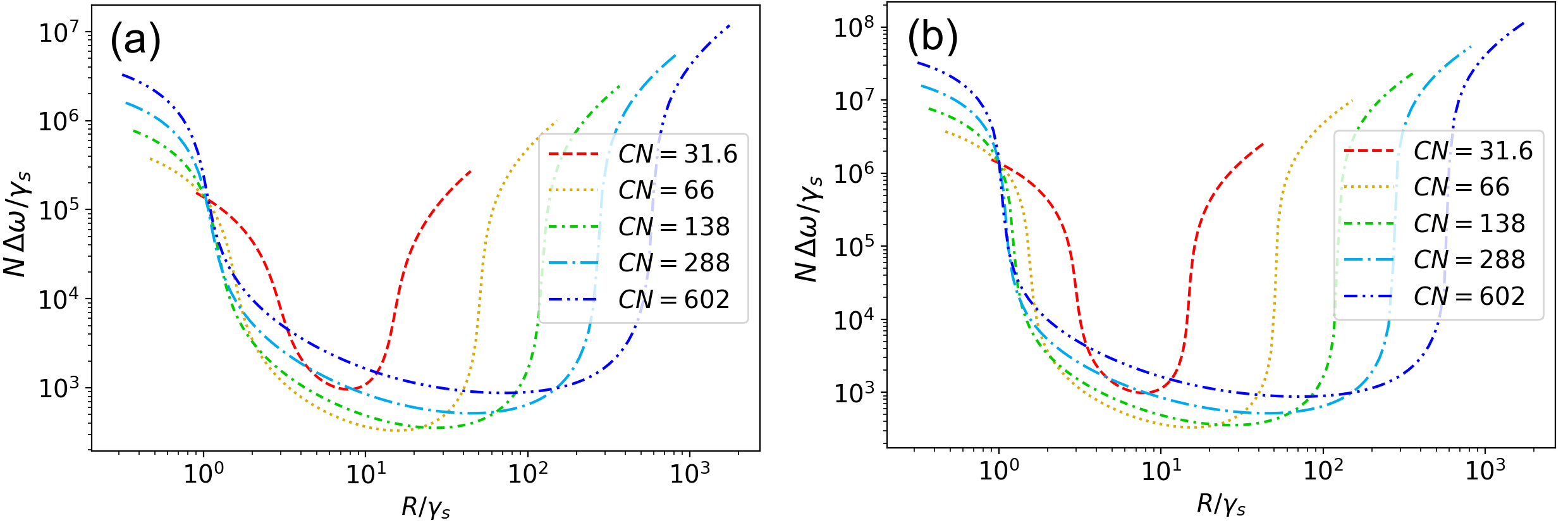}
    \caption{Dependency of linewidth $\Delta \omega$ on repumping rate $R$ for a homogeneous system at different values of $CN$ for different values of number $N$ of atoms and finesse $\FF$ of the cavity. (a): $N=10^4$, $\FF=10^4$. (b): $N=10^5$, $\FF=10^5$. In both cases the atomic dephasing rate is $\gamma'=0.1~{\rm s^{-1}}$ and the cavity length is $\lcav=10~$cm, which corresponds to $\kappa=\pi c/(\FF \lcav)\approx9.4\times 10^5~{\rm s^{-1}}$ and $\kappa\approx9.4\times 10^4~{\rm s^{-1}}$ respectively.}
    \label{fig:FWHMhom}
\end{figure}
\subsection{Minimized linewidth}

\label{Sec::modelCalc}
In this subsection we investigate in more details the dependence of the optimized spectral linewidth $\Delta \omega$ on various parameters of the superradiance laser. 
First, we consider the homogeneous case. In Figure~\ref{fig:FWHMhom} we present the linewidth $\Delta \omega$ for different values of $CN$ as function of incoherent repumping rate $R$, calculated according to the method described in subsection~\ref{Sec::modelHomogen}. 
One may see that, being expressed in units of $\gamma_s/N$, all the linewidths show a quite similar behaviour, except near the lower and the upper lasing thresholds.
\begin{figure}
    \centering
    \includegraphics[width=0.85\textwidth]{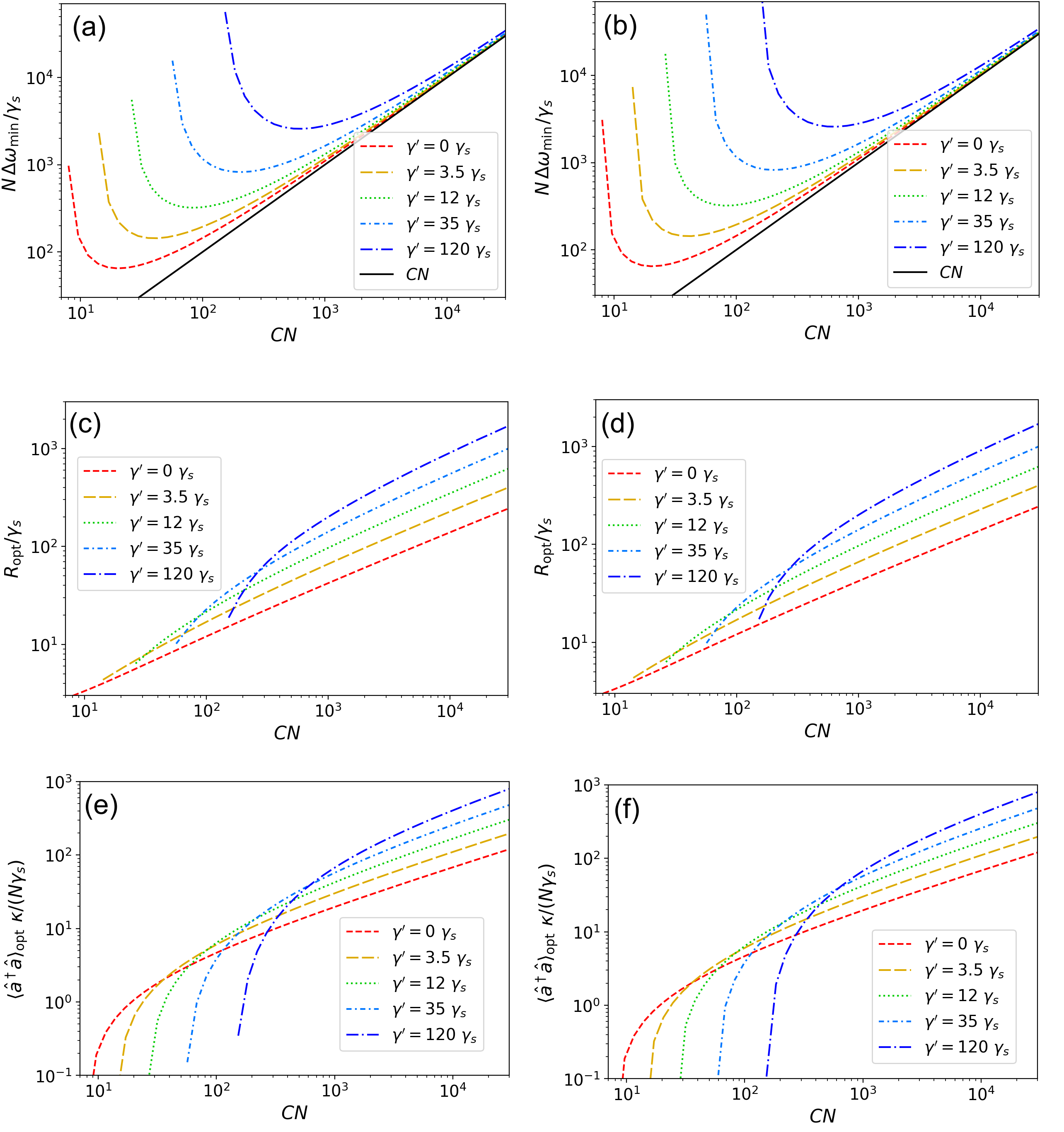}
    \caption{Dependency of minimum attainable linewidth $\Delta \omega_\text{min}$,  
    the optimal repumping rate $R_\text{opt}$ in units of $\gamma_s$ ((c),(d)), 
    and the respective intracavity photon number $\langle \hat{a}^\dagger \hat{a}\rangle_\text{opt}$ multiplied by $\kappa/(N\gamma_s)$ ((e),(f))) 
    on the parameter $CN$ for different values of atomic dephasing $\gamma'$.
    The graphs are for different values of number $N$ of atoms and finesse $\FF$ of the cavity: 
    (a, c, e): $N=10^4$, $\FF=10^4$. (b, d, f): $N=10^5$, $\FF=10^5$. In (a) and (b) the asymptotic $CN$ behavior is indicated by a black line.
    }
    \label{fig:FWHMhomOpt}
\end{figure}

For any of the curves, similar to the ones presented in Figure~\ref{fig:FWHMhomOpt}, we can find the minimum linewidth $\Delta \omega_\text{min}$, obtained at some optimal repumping rate $R_\text{opt}$. 
In Figure~\ref{fig:FWHMhomOpt} we present the dependency of these minimized linewidths on $CN$ for different values of the atomic dephasing rate $\gamma'$, number $N$ of the atoms and the cavity finesse $\FF$. 
Note that the value $\Delta \omega_{\rm min}$ expressed in units $\gamma_s/N$ as well as the optimal repumping rate $R_\text{opt}$
does not depend on $N$ (i.e. the optimized linewidth $\Delta\omega_\text{min}$ is inversely proportional to $N$ at a given value of $CN$). 
Similarly, the ratio of $\langle \ad \a \rangle \cdot \kappa$ to $N\gamma_s$ corresponding to the minimized linewidth as well as the optimal repumping rate $R_\text{opt}$ depend on the atomic dephasing rate $\gamma'$ but not on $\FF$ or $N$. 
In this example the cavity length $\lcav$ has been taken as $\lcav=10~$cm, although the results are not sensitive to variations of the cavity length as long as the laser operates in the bad-cavity regime, as discussed in section~\ref{sec:estimation}.

We should also note that the value $\langle \ad \a \rangle \cdot \kappa/(N\gamma_s)$ has a simple physical interpretation: it is the ratio of number of photons emitted from the cavity mode (in case of perfect outcoupling mirror $\eta=1$) to the single-atom spontaneous emission rate $\gamma_s$ multiplied by the number of atoms. Near the maximum of the output power this ratio is proportional to $N$, however, near the minimum of the linewidth it is independent on $N$. In the absence of atomic dephasing, the minimum attainable linewidth (optimized by both the repumping rate $R$ and the cooperativity $C$) is about $\Delta\omega_{opt} \approx 64 \gamma_s/N$.

Up to now we calculated the linewidths in the frame of a fully homogeneous model. 
However, in real systems different atoms may expect different level shifts, they may expect different dephasings due to interaction with environment, and different pumping rates. 
Last but not least, different atoms can be coupled differently with the superradiance cavity field. This happens particularly when the atoms trapped within the magic optical lattice created inside the superradiance cavity are coupled to the standing-wave mode of the same cavity, because of the mismatch of the magic wavelength trapping the atoms and the wavelength of the superradiance mode, see expression~\ref{eq:43} in section~\ref{sec:estimation}. The spectral linewidth of the superradiance radiation can be calculated using the method described in subsection~\ref{Sec::modelInhomogen}.
\begin{figure}
    \centering
    \includegraphics[width=0.85\textwidth]{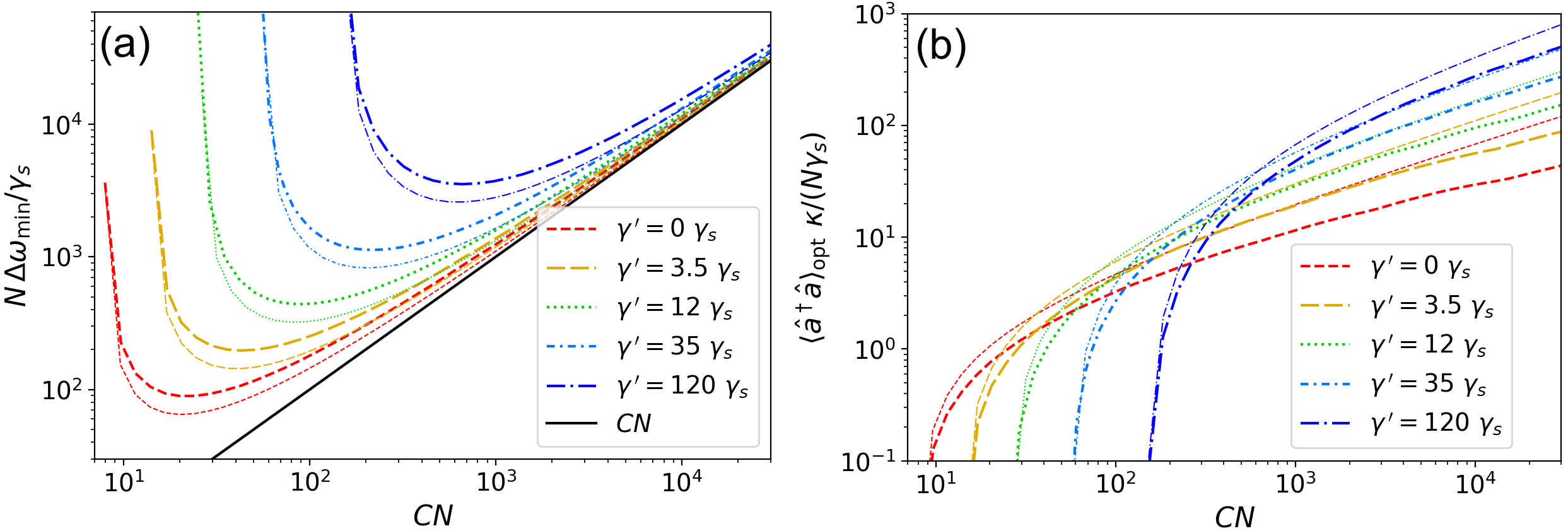}
    \caption{Dependency of minimum attainable linewidth $\Delta \omega_\text{min}$ in units of $\gamma_s/N$ (a), and the respective intracavity photon number $\langle \hat{a}^\dagger \hat{a}\rangle_\text{opt}$ multiplied by $\kappa/(N\gamma_s)$ (b) on the parameter $CN$ for a system with inhomogeneous cosine-modulated coupling (thick curves) system for different values of atomic dephasing rate $\gamma'$ at $N=10^5$, $\FF=10^5$. 
    The cavity length is $\lcav=10~$cm.
    Thin curves represent the linewidths and the intracavity photon numbers calculated according to the homogeneous model, the same colour and style corresponds to the same value of $\gamma'$.}
    \label{fig:FWHMInhomOpt}
\end{figure}
In Figure~\ref{fig:FWHMInhomOpt} we present the dependencies of the minimum attainable linewidth $\Delta \omega_\text{min}$ and the intracavity photon number $\langle \ad \a \rangle$ on cooperativity $CN$, calculated for repumping rates $R_\text{opt}$ which minimise the linewidth. 
We grouped the atoms into $M=21$ cluster containing equal numbers of atoms.
Coupling coefficients $g_j$ for $j$th cluster were taken proportional to $\cos(\frac{\pi (j-0.5)}{2 M})$; all the other parameters are the same for all the clusters, also $\Delta_j=\delta_c=\xi=0$. 
The cooperativity $C$ is defined according to
\begin{equation}
CN=\sum_j \frac{4 g_j^2}{\kappa \gamma_s}.
\label{eq:37}
\end{equation}
 
For comparison, we present the dependencies of $\Delta\omega_\text{min}$ and $\langle \ad \a \rangle_\text{opt}$ calculated according to the homogeneous model. 
One can see that the homogeneous model slightly underestimates the attainable linewidth and overestimates the intracavity photon number, both by a factor of about 1.4 near the optimally chosen $CN$. 
Particularly, at $\gamma'=12\gamma_s$ the minimum linewidth s $\Delta \omega \approx 4.3\times 10^2~\gamma_s/N$ for inhomogeneous coupling, and $\Delta \omega \approx 3.1\times 10^2~\gamma_s/N$ for homogeneous coupling.

\begin{figure}
    \centering
    \includegraphics[width=0.85\textwidth]{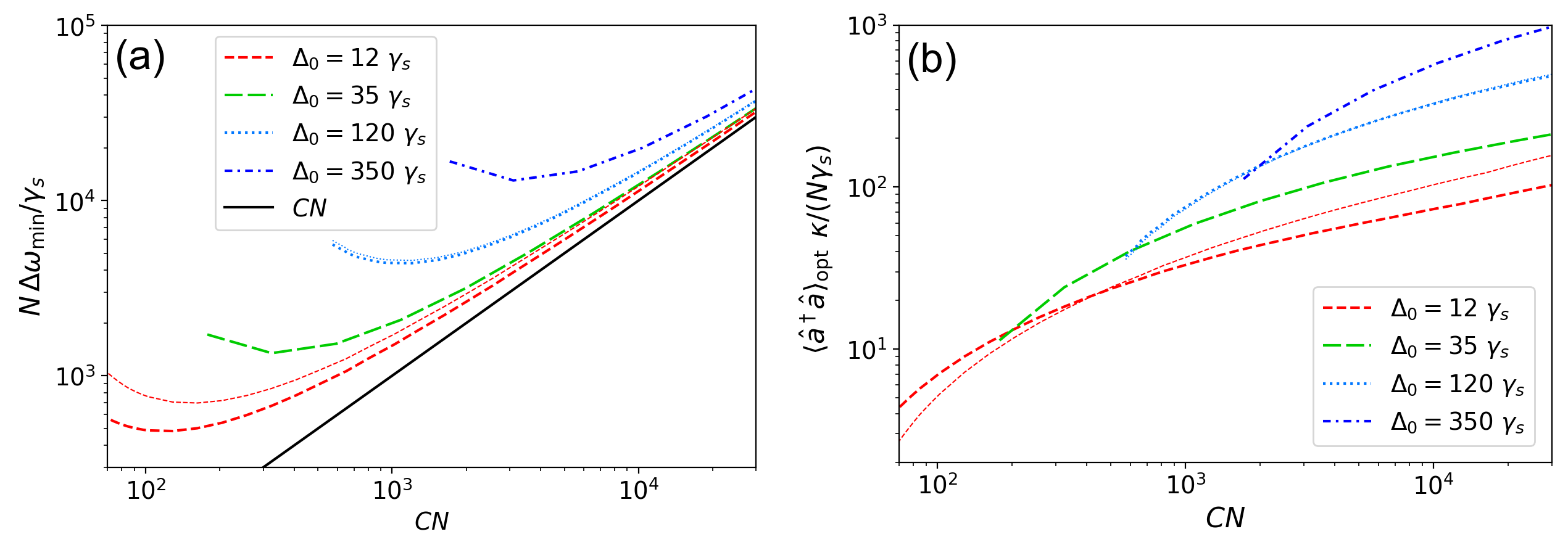}
    \caption{Dependency of minimum attainable linewidth $\Delta \omega_\text{min}$ in units of $\gamma_s/N$ (a), and the respective intracavity photon number $\langle \hat{a}^\dagger \hat{a}\rangle_\text{opt}$ multiplied by $\kappa/(N\gamma_s)$ (b) on the parameter $CN$ for a system with inhomogeneous cosine-modulated coupling for different values of broadening $\Delta_0$ at $N=10^5$, $\FF=10^5$. 
    The atomic dephasing rate is $\gamma'=0$ (thick curves), and $\gamma'=12\gamma_{s}$ (thin curves; only for $\Delta_0=12\gamma_s$ and $\Delta_0=120\gamma_s$, the same color-style encoding corresponds to the same values of $\Delta_0$). The cavity length is $\lcav=10~$cm.}
    \label{fig:FWHMInhomBroadOpt}
\end{figure}

Figure~\ref{fig:FWHMInhomBroadOpt} shows the minimised linewidth $\Delta\omega_\text{min}$ for a system where not only the coupling of the atoms to the cavity mode is inhomogeneous, but also the lasing transitions in different atoms have different shifts $\Delta_j$. 
Such shifts can be caused by variations of environmental parameters over the atomic ensemble. 
Here we considered the simplest case where the atomic detunings $\Delta_j$ are evenly distributed over 11 clusters between $\pm \Delta_0$, and the couplings are also distributed over 7 clusters; therefore we have 77 clusters in total. 
At 
$\gamma'=\Delta_0=12\gamma_s$ the minimum attainable linewidth $\Delta \omega_\text{min} \approx 7\times 10^2~\gamma_s/N$, whereas increasing $\Delta_0$ to 
$120~\gamma_s$ would increase the linewidth to about 
$\Delta \omega_\text{min} \approx 4.65\times 10^3~\gamma_s/N$.

Finally, it is useful to consider the dependence of the linewidth $\Delta \omega_\text{opt}$ - doubly minimized both in $R$ and $CN$ - on the dephasing rate $\gamma'$ and on the inhomogeneous broadening $\Delta_0$. By fitting the result of the simulations we obtain the estimated linewidth in the form
\begin{equation}
\Delta \omega_\text{opt} \approx (90 \gamma_s+30 \gamma' + 35 \Delta_0)/N.
\label{eq:38}
\end{equation}
Expressing the linewidth via the more useful dispersion of the shifts $\Delta^\prime_0 = \Delta_0/\sqrt{3}$ for the flat distribution assumed in the simulations, gives approximately
\begin{equation}
\Delta \omega_\text{opt} \approx (90 \gamma_s+30 \gamma' + 60 \Delta^\prime_0)/N
\label{eq:39}
\end{equation}

Similarly one can find approximate expressions for the optimal pumping rate $R_{\rm opt}$, for the collective cooperativity $CN_{\rm opt}$, and for the intracavity photon number, where the smallest linewidth $\Delta\omega_{\rm opt}$ is achieved: 

\begin{align}
R_{\rm opt} & \approx 5 \gamma_s+1.13 \gamma' + 1.5 \Delta^\prime_0 
\label{eq:40} \\
CN_{\rm opt} & \approx 25 
         + 5.5 \frac{\gamma'}{\gamma_s}
         + 20\frac{\Delta^\prime_0}{\gamma_s} 
\label{eq:41} \\
\langle \ad \a \rangle_{\rm opt} & \approx \frac{N}{\kappa}
    \left(
        0.9\gamma_s + 0.25 \gamma'+1.45 \Delta^\prime_0 
    \right)
\label{eq:42}
\end{align}

\section{Estimation of attainable stability}
\label{sec:estimation}

To perform quantitative estimations, we need to consider realistic parameters of the atomic ensemble. The double forbidden ${^1S_0}\leftrightarrow {^3P_0}$ transition (clock transition) in fermionic isotopes of alkaline-earth-like atoms (Be, Mg, Ca, Sr, Zn, Cd, Hg and Yb) seems to be a good choice for optical clocks with neutral atoms. 
This transition is totally forbidden in bosonic isotopes and becomes slightly allowed in fermionic isotopes by hyperfine mixing. 
These atoms can be trapped in a the magic-wavelength optical lattice potential and pumped into the upper ${^3P_0}$ lasing state. 
In an active optical clock the clock transition should be coupled to a high-finesse cavity in the strong cooperative coupling regime, which is problematic for wavelengths of about 458~nm (corresponding to clock transition in Mg) and shorter. 
Therefore, Ca, Sr and Yb with wavelengths of the clock transition $\lambda$ equal to $660$, $698$ and $578$~nm, respectively, are the most feasible candidates for the role of gain atoms in active optical clocks. 
In the present paper we will primarily perform our estimations for the ${\rm ^{87}Sr}$ isotope, because, first, this element is the most used one in modern optical clocks with neutral atoms and its relevant characteristics are the most studied among all the alkaline-earth-like atoms. 
Second, the natural linewidth of the clock transition in ${\rm ^{87}Sr}$ ($\rm \gamma_s = 8.48\times 10^{-3}~s^{-1}$ \cite{Muniz2020}) lies between the linewidths of $^{43}\text{Ca}$ ($\rm 2.2\times 10^{-3}~s^{-1}$) and Yb ($\rm 43.5\times 10^{-3}~s^{-1}$ and $38.5\times 10^{-3}~\mathrm{s^{-1}}$ for ${\rm ^{171}Yb}$ and ${\rm ^{173}Yb}$ respectively) \cite{Porsev04}.

The finesse $\FF$ of the best cavities at a wavelength of 698~nm can reach values of up to $10^6$, however, it is quite difficult to build such a cavity. 
More feasible finesse values would range from tens to hundreds of thousands. 
For the sake of definiteness, we take $\FF=10^5$ as a typical parameter. 

The coupling strengths $g_j$ between the lasing transition in the $j$th atom and the cavity field can be estimated as
\begin{equation}
g_j\approx \frac{1}{w_c}\sqrt{\frac{6 c^3 \gamma_s}{\lcav \omega_0^2}} \cos(k_0 z_j),
\label{eq:43}
\end{equation}
where $k_0=\omega_0/c$ is the wave number of the cavity mode, $w_c$ is the cavity waist radius, and $z_j$ is the $z$-coordinate of the $j$th atom along the cavity axis \cite{Gogyan20}. 
For the sake of simplicity, here we neglect the dependency of the coupling strength $g$ on the distance from the atom to the cavity axis proportional to $\exp(-(x_j^2+y_j^2)/w_c^2)$ (which can be relevant for atoms trapped in 2D or 3D optical lattices as well as for relatively hot atomic ensembles in shallow 1D optical lattice). 
Note that the cooperativity $C=4\sum_{j} g_j^2/(N\kappa \gamma_s)$ does not depend on the length of the cavity $\lcav$ but only on the cavity finesse $\FF$ and the cavity mode waist $w_c$, because both $g_j^2$ and $\kappa$ are inversely proportional to $\lcav$. 
Therefore, the cavity length $\lcav$ is not a very important parameter, as long as the energy decay rate $\kappa = \pi c/(\lcav \FF)$ of the cavity mode is much larger than the linewidth of the laser gain. 
For the calculations performed in section~\ref{sec:linewidth} we take $\lcav=10~\text{cm}$, which corresponds to a decay rate $\kappa = 9.42 \times 10^4 ~\mathrm{s}^{-1} \approx 2\pi \times 15~{\rm Hz}$ at $\FF=10^5$.

Let us first compare the ultimate stability of an incoherently pumped active optical frequency standard with the stability of a quantum projection noise (QPN) limited passive frequency standard, assuming the same number of trapped atoms in both standards and no inhomogeneous broadening or decoherence. 
The fundamental limit of the superradiant laser linewidth is then $\Delta \omega \approx 90~\gamma_s/N$, as follows from expression~(\ref{eq:39}).
This corresponds to a short-term stability
\begin{equation}
\sigma_{y,\text{lim}} (\tau) \approx \frac{1}{\omega}\sqrt{\frac{90 \gamma_s}{N \tau}} \approx \frac{9.5}{\omega} \sqrt{\frac{\gamma_s}{N \tau}}.
\label{eq:44}
\end{equation}
For passive optical clocks the quantum projection noise limited stability $\sigma_{y,\text{QPN,Rams}}$ and $\sigma_{y,\text{QPN,Rabi}}$ for Ramsey and Rabi interrogation schemes respectively can be estimated as \cite{Ludlow_2015, Oelker_2019}
\begin{align}
\sigma_{y,\text{QPN,Rams}} (\tau) 
    &= \frac{1}{\omega \sqrt{N T_p \tau}}
\label{eq:45}, \\
\sigma_{y,\text{QPN,Rabi}} (\tau) 
    &\approx \frac{1.69}{\omega \sqrt{N T_p \tau}} \label{eq:46},
\end{align}
if the total Rabi or Ramsey interrogation time $T_p$ is much longer than all the other durations required for state preparation and measurement, 
and if it is much shorter than the excited state lifetime $1/\gamma_s$.
Comparing equations (\ref{eq:44}) with (\ref{eq:45}) and (\ref{eq:46}) one may see that at the same atom number the ultimate stability (\ref{eq:44}) attainable with an active optical clock with incoherent pumping can be matched by the QPN limited stability of a passive clock, at interrogation times of $T_p=1/(90\,\gamma_s)\approx 0.011/\gamma_s$ for Ramsey, and at $T_p=1.69^2/(90\,\gamma_s) \approx 0.032/\gamma_s$ for Rabi interrogation. 
For clocks using ${\rm ^{87}Sr}$ these times are $T_p=1.31~{\rm s}$ for Ramsey, and $T_p=3.74~{\rm s}$ for Rabi interrogation. 
For the ${^1S_0}\leftrightarrow{^3P_0}$ transition in ${\rm ^{173}Yb}$ the corresponding times are 0.25~s and 0.72~s respectively, and for ${\rm ^{43}Ca}$ 5.05~s and 14.4~s .

A more realistic comparison between the achievable stability of the active and passive optical frequency standards must include additionally dephasing of the atomic transition, as well as imperfections of the local oscillator in a passive clock . 
The transverse dephasing rate $\gamma'=2/T_2$ of the atomic transition is limited by Raman scattering of photons from the optical lattice potential \cite{doe18}, and by site-to-site tunneling of the atoms \cite{Lemonde05}. 
In a shallow cubic 3D optical lattice with $^{87}$Sr \cite{Hutson19} an optimized coherence time $T_2 \approx 10~\text{s}$ was achieved, which corresponds to $\gamma' \approx 0.2~\text{s}^{-1}$.
This decoherence time may be even further reduced with the help of technically more challenging setups , such as using of optical lattices with increased lattice constants formed, for example, by interfering laser beams at different angles or by optical tweezer arrays
\cite{Hutson19} .
Moreover, collisions with residual background gas also destroy the coherence and reduce the trap lifetime. 
From this point of view, $\gamma'=0.2~\text{s}^{-1}$ seems to be a good estimate for the minimum atomic decoherence rate that can be achieved without extraordinary efforts.
Assuming an inhomogeneous broadening $\Delta_0$ of the atomic ensemble of $\Delta_0 \approx 2\pi \times 15~{\rm mHz} \approx 0.09~{\rm s^{-1}}$, 
one may estimate the optimized linewidth $\Delta \omega_\text{opt}$ of the superradiance laser as 
$\Delta \omega_\text{opt} \approx 10/N~\text{s}^{-1}$, corresponding to a stability of a $^{87}$Sr active clock 
\begin{equation}
	\sigma^\prime_y(\tau) =\frac{1}{\omega}\sqrt{ \frac{\Delta \omega}{\tau}} \approx \frac{1.17\times 10^{-15}}{\sqrt{N \, \tau }} .
	\label{eq:47}
\end{equation}
For $N = 10^4$ it results in an instability of $10^{-17}$ at 1~s of averaging, and of $10^{-18}$ after 100 seconds, whereas a bad-cavity laser with $N=10^5$ atoms would provide an instability of
$\sigma^\prime_y(\tau) \approx 3.7 \times 10^{-18}/\sqrt{\tau[\text{s}]}$. 

Let us now compare this stability with the one that can be attained in a passive clock with the same number of atoms. 
An ideal quantum projection noise-limited, zero dead time, passive $^{87}$Sr optical clock can attain such a stability at interrogation times of $T_p=0.1~{\rm s}$ for Ramsey, and $T_p=0.29~{\rm s}$ for Rabi interrogation, as follows from equations (\ref{eq:45}) and (\ref{eq:46}). 
These interrogation times are short compared to the inverse inhomogeneous broadening and to the decoherence time of the atomic ensemble as estimated above,
thus, these effects would not yet limit the passive clock.
However, in a passive optical clock based on the sequential discontinuous interrogation of the clock transition in single atomic ensembles, the frequency fluctuations of the local oscillator contribute substantially to the instability due to the Dick effect \cite{dic87}. 

\begin{figure}[htb]
    \centering
    \includegraphics[width=0.6\textwidth]{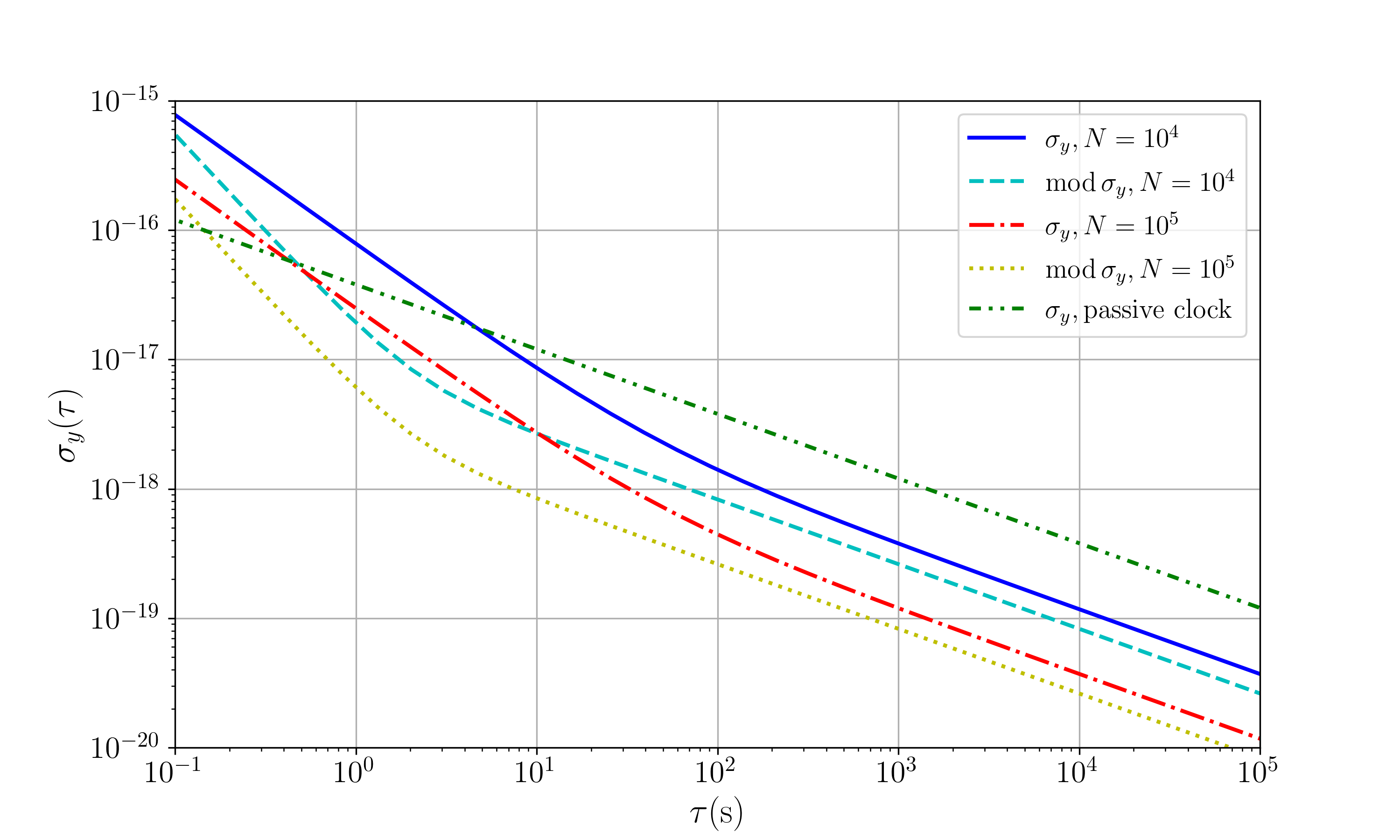}
    \caption{Stability of the ${}^{87}$Sr active clock output expressed as Allan deviation $\sigma_y$ with $f_h = 10~$Hz for $N=10^4$ (blue solid line) and $N=10^5$ (red dash-dotteded line). 
    The corresponding modified Allan deviation $\mathrm{mod}~\sigma_y$ is shown by the cyan dotted line and the yellow dotted line. 
    The different slopes are due to contributions from photon shot noise and atomic phase diffusion. 
    For comparison, the stability of a Dick effect limited passive clock as discussed in the text is shown as green dash-dot-dot line.}
    \label{fig:ADEV}
\end{figure}

For example, in ref. \cite{Oelker_2019} the contribution to instability $\sigma_{y,\text{Dick}}$ from this Dick effect was on the level of $\sigma_{y,\text{Dick}}\approx 3.8 \times 10^{-17}/\sqrt{\tau[\text{s}]}$ (see Fig. \ref{fig:ADEV}). 
Such a level of stability has been obtained with a local oscillator laser pre-stabilized to an elaborate 21~cm cryogenic silicon resonator at 124~K. 
The bad cavity laser can provide similar stability at a linewidth $\Delta \omega \approx 0.01~\text{s}^{-1}$, {that can } be attained with $N=10^4$ atoms and a dephasing rate $\gamma' \approx 1.5~\text{s}^{-1}$, or with $N=10^5,~\gamma' \approx 5~\text{s}^{-1}$ ($T_2 = 2/ \gamma'=0.4~\text{s}$), if the inhomogeneous broadening $\Delta^\prime_0$ is much less than the dephasing rate. 
Therefore, the short-term stability of {an active optical frequency standard may match and even significantly exceed the stability of passive clocks limited to the noise of local oscillator via the Dick effect.
On the other hand, the quantum projection noise-limited stability of a passive clock based on a similar atomic ensemble can be still better than the one of the passive standard.}

We should note, that the Dick effect in passive optical clocks can be avoided (or at least significantly suppressed down to contributions of finite-length $\pi/2$ pulses) by an interleaved, zero dead time operation of two clocks \cite{Schioppo17}. 
When comparing two clocks using the same atomic transition, the Dick effect can also be eliminated and the interrogation time extended to beyond the coherence time of the laser by using synchronous interrogation \cite{Schioppo17, Takano16, Oelker_2019} of the two atomic ensembles. 
In the extreme case, comparing different parts of the same cloud, a fractional instability of 
$\sigma_{y}\approx 4 \times 10^{-18}/\sqrt{\tau[\text{s}]}$ could been achieved \cite{bot22}.
Similarly, comparing clocks on operating on different atomic transitions, differential spectroscopy \cite{Kim21} or dynamical decoupling methods \cite{doe20} can be employed. 

\begin{figure}[htb]
    \centering
    \includegraphics[width=0.6\textwidth]{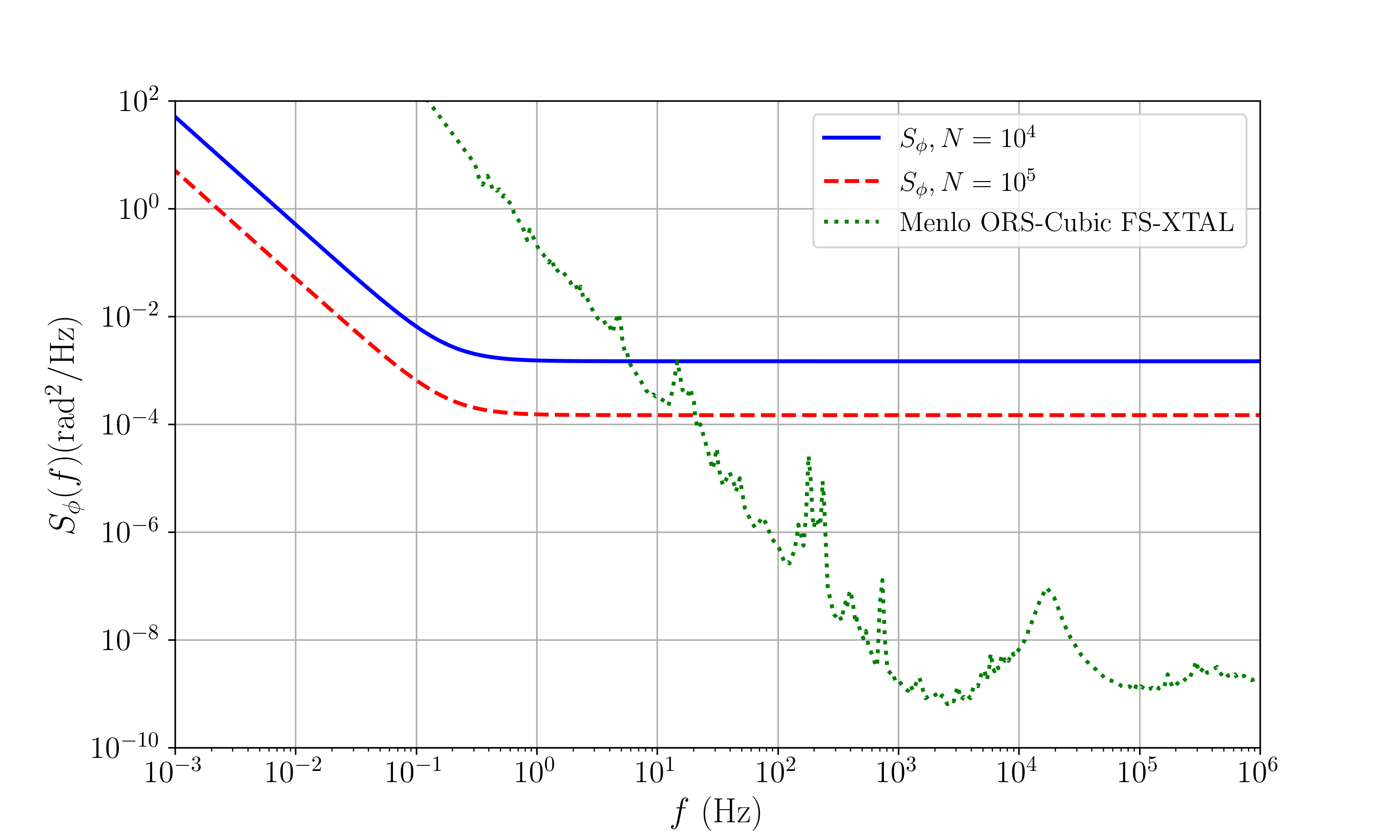}
    \caption{Spectral power density of phase fluctuations $S_\phi$ for an active clock with $10^4$ atoms (blue line), $10^5$ atoms (dashed red line) and for a commercial cavity stabilized laser (green dash-dotted line)
    }
    \label{fig:Sphi}
\end{figure}

At the optimum stability the output power $P$ of the bad cavity laser amounts to a photon flux of
$P/\hbar \omega_0 = \eta \kappa \langle \hat{a}^\dagger \hat{a} \rangle \approx \eta N(0.9\gamma_s+0.25 \gamma'+1.45 \Delta_0^\prime)$, 
see expressions~(\ref{eq:20}) and (\ref{eq:42}). 
Taking $\eta=0.5$ and parameters of the atomic ensemble listed above ($\gamma_s=8.48 \times 10^{-3}~{\rm s^{=1}}$, $\gamma'=0.2~{\rm s^{=1}}$ and $\Delta_0^\prime=\Delta_0/\sqrt{3}\approx 0.054~{\rm s^{-1}}$), the photon flux at the optimized cooperativity and pumping rate will be about $680~\mathrm{s}^{-1}$ for $N=10^{4}$ and $6800~\mathrm{s}^{-1}$ for $N=10^{5}$.  

This output power of the active clock is usually too small for practical application, thus a suitable secondary laser needs to be phase locked to the weak output to boost the available power. 
The bandwidth of this phase-lock depends on the stability of the (shot noise limited) active clock and the stability of the free running secondary laser. 
In Fig. \ref{fig:Sphi} the phase noise of the superradiant laser output for $10^4$ and $10^5$ atoms is shown in comparison to the phase noise of a commercial laser system, based on a 5 cm long cubic cavity (Menlo Systems OFR-cubic FS-XTAL \cite{MenloCubic}) with fractional frequency instability $\text{mod}\, \sigma \approx 10^{-15}$ .
For best overall performance, the bandwidth should extend up to the crossing between the phase noise curves of the secondary laser and of the superradiant laser. 
In the example shown here this crossing is around 10 Hz in agreement with our previous choice of a 10 Hz cut-off frequency $f_h$ for the white phase noise contribution to the Allan deviation. 

The Allan deviations for the superradiant laser output are shown in Fig. \ref{fig:ADEV}. Including the phase locked laser would only cap the strong increase of the stability towards short averaging times and limit the instability to values of $10^{-15}$ below 0.1~s. 

Besides the fundamental limit to the stability from the superradiant laser's linewidth, also the
 stability of the active clock may degrade due to a drift or fluctuations of the environmental parameters, such as the bias magnetic field. 
For example, the Zeeman shift of the $\pi$-transition $|^3P_0,m\rangle \rightarrow |^1S_0,m\rangle$ in ${\rm ^{87}Sr}$ amounts to about 
$\Delta \omega / B = \mathrm{2\pi \cdot 1.10~Hz / \mu T} \cdot m_F $
\cite{Boyd_2007}, that results in a shift of about $2\pi \times 4.95~{\rm Hz/\mu T}$ for transition between the two stretched states $|^3P_0,m=9/2\rangle$ and $|^1S_0,m=9/2\rangle$. 
For example, to attain a $10^{-18}$ level of relative uncertainty of the clock transition frequency, one has to decrease the uncertainty of the bias magnetic field to below $87~{\rm pT}$. 
In passive clocks the linear Zeeman effect is usually canceled by taking the average between Zeeman transitions with opposite shifts alternating from one interrogation cycle to the other. 
This method eliminates drifts and slow fluctuations of the bias magnetic field but can not cancel fluctuations on timescale below a single interrogation cycle duration. 
In contrast, active clocks may operate on two Zeeman transitions simultaneously, 
 generating two-frequency laser radiation from both $\pi$-transitions between pairs of stretched states $|^3P_0,m=9/2\rangle \rightarrow |^1S_0,m=9/2\rangle$ and $|^3P_0,m=-9/2\rangle \rightarrow |^1S_0,m=-9/2\rangle$. 
The arithmetic mean of both these frequencies will be robust to fluctuations of the first-order Zeeman shift, as well as of a vector Stark shift from the lattice field. 
Both transitions can contribute independently to lasing, if they both interact with the same mode of the cavity and if they are detuned from each over far enough to neither get synchronized nor significantly affect each other. 
This condition can be easily attained under realistic conditions: for example, a bias magnetic field $B=1~{\rm G}=0.1~{\rm mT}$ splits these two transitions by about $2\pi \times 1$~kHz. 
This splitting is less than the linewidth $\kappa$ of the cavity (estimated above as $\kappa \approx 2\pi \times 15~{\rm kHz}$ at $\lcav=10~{\rm cm}$ and $\FF=10^5$), but it is much larger than the optimized pumping rate 
$R_\mathrm{opt} \approx 0.35~\mathrm{s}^{-1}$ 
\cite{Kazakov2017, Xu13}, as estimated from equation (\ref{eq:40}).

\section{Conclusion}
\label{sec:concl}
In this paper we studied the ultimate frequency stability that can be obtained with active optical frequency standards. 
We investigated the dependence of the linewidth of a bad-cavity laser with incoherent pumping on its parameters and obtained an estimated minimum linewidth (Eq. \ref{eq:38}) under optimized conditions. 
We showed that the instability $\sigma_{y,\text{Dick}}\approx 3.8 \times 10^{-17}/\sqrt{\tau[\text{s}]}$ of a passive optical frequency standard associated with the Dick effect for one of the best local oscillators pre-stabilized to a cryogenic Si cavity \cite{Oelker_2019} can be matched by a bad-cavity laser with $N=10^5$ ${^{87}\text{Sr}}$ atoms with coherence time $T_2\approx 0.4~{\text{s}}$.
As active optical frequency standards are not degraded by the Dick effect associated with dead time and noises of the local oscillator, they can outperform ``traditional'' passive optical frequency standards in stability.
Also, active optical frequency standard may play a role as local oscillators in future passive optical clocks. 
Even if their short-term stability is poorer by small factor than the quantum projection noise limited stability of a passive optical clock with a similar number of clock atoms, the stability can still be significantly better than that of a good-cavity laser pre-stabilized to an ultra-stable cavity, as used in modern passive optical clocks.

\section*{Acknowledgment}
We acknowledge support by Project 17FUN03 USOQS, which has received funding from the EMPIR programme cofinanced by the Participating States and from the European Union’s Horizon 2020 Research and Innovation Programme,
by the European Union Horizon 2020 research and innovation programme Quantum Flagship projects No 820404 ``iqClock'', No 860579 ``MoSaiQC'', 
Narodowe Centrum Nauki (Quantera Q-Clocks 2017/25/Z/ST2/03021) and SFB 1227 DQ-mat, Project-ID 274200144, within Project
B02.

Numerical simulations were performed with the open source frameworks DifferentialEquations.jl~\cite{Rackauckas17}. 
The graphs were produced using the open source plotting library Matplotlib~\cite{hunter2007matplotlib}. 
Programs to simulate physical models are available at Zenodo repository, \url{https://zenodo.org/record/6500087}
\bibliography{main}

\end{document}